\documentclass[aps,pra,superscriptaddress,amsmath,amssymb,preprintnumbers,floatfix,showpacs,12pt]{revtex4}
\usepackage{amssymb}
\usepackage{epsfig}
\usepackage{subfigure}
\usepackage{graphicx}
\begin{document}
\title{ Generation of squeezed light in a nonlinear asymmetric directional
coupler  }

\author{Faisal A. A. El-Orany\footnote{Permanent address: Suez Canal university, Faculty of
Science, Department of mathematics and computer science, Ismailia,
Egypt.}, J. Pe\v{r}ina  } \affiliation{ Joint Laboratory of Optics
of Palack\'y University and Physical Institute of Academy of
Sciences of Czech Republic, 17.~listopadu~50, 772 07~Olomouc,
Czech Republic. }

\author{M. Sebawe Abdalla}

\affiliation{Mathematics Department, College of Science, King Saud
University, P.O. Box 2455, Riyadh 11451, Saudi Arabia}

\begin{abstract}
We show that a nonlinear asymmetric directional coupler composed
of a linear waveguide and a nonlinear waveguide operating by
nondegenerate parametric amplification is an effective source of
single-mode squeezed light.
 This is has been demonstrated, under certain conditions and for specific
 modes, for incident coherent beams in terms of the quasiprobability
 functions, photon-number distribution and phase distribution.

\end{abstract}
 \maketitle


\section{ Introduction}

Squeezed states of light offer possibilities of improving the performance
of optical devices since they can reduce fluctuations in one quadrature
below the  level associated with the vacuum states. Such a situation is relevant
for the optical communication networks as well as for many optical devices.
For instance, recently such light  has been used
in a power recycled interferometer \cite{int1} and in a
phase-modulated signal-recycled interferometer \cite{int2} aiming to
 improving significantly the sensitivity of these
devices. It has been shown that  this light can be used to
tune the resonant frequency of the cavity without actually moving
the signal recycling mirror or  changing the bandwidth of the
interferometer without substantially decreasing the sensitivity at
the resonant frequency \cite{int2}.
Also squeezed light has been applied in quantum information theory, e.g.
in a quantum teleportation \cite{inf1},
   cryptography \cite{inf2,{inf3}} and dense coding \cite{inf4}.
In this respect the security in a quantum cryptography \cite{inf3}   relies
on the uncertainty relation for field quadrature components of these
states.  Further, experiments on quantum teleportation have been successfully
performed by means of two-mode squeezed vacuum states \cite{tele}.
A variety of methods have been proposed for the generation of squeezed states
of the electromagnetic field and several of them have been realized
experimentally \cite{ex1,{ex2}}.
 For more complete information about squeezed
light the reader can consult review papers \cite{[23]}.

As is well known the ideal single-mode  squeezed light (with
possible $100\%$ squeezing) may be generated using a degenerate
parametric amplifier (DPA) with classical pump. This is the most
effective system in nonlinear optics in which such strong
squeezing was obtained. DPA is the two-photon  process  where the
signal and idler modes are identical in this case. Such a property
manifests \cite{fai} itself in the large scale oscillations in the
photon-number distribution \cite{sh} and in the bifurcation in
phase distribution \cite{mir1}. In other words, the occurrence of
squeezing in the quadrature variances does not need to be
accompanied by  oscillations in the photon-number distribution or
bifurcation in the phase distribution. The best example  are the
binomial states \cite{kni}. On the other hand, the nondegenerate
parametric amplifier (NDPA) can produce ideal squeezing in the
compound modes even if the single mode squeezing does not appear
\cite{fais}.

In this article we suggest a compound system--nonlinear three-mode asymmetric
directional coupler  which is able to provide
 both strong single mode squeezing  and compound two mode squeezing
under certain conditions and for specific modes.
This means that the present system can operate simultaneously as degenerate and nondegenerate
amplifiers.
Moreover,  the single mode squeezing can  be confirmed in
the behaviour of both the  photon-number distribution and  phase
distribution.
In general, the nonlinear directional coupler consists
of two or more parallel optical waveguides fabricated from  nonlinear
material. Both waveguides are placed close enough to permit flux-dependent
transfer of energy between them as a consequence of evanescent waves
involved in the interaction \cite{marc}. This flux transfer can be
controlled by the device design and the input flux.
Further, the nonlinear directional coupler has been experimentally implemented
, e.g.  using semiconductor/glass composite
\cite{Li1}, erbium doped fibers  \cite{jin1,{gus}} and a crystalline organic
semiconductor \cite{Tow}.
Furthermore, the directional couplers  are of the technological interest since
they are currently a topic of major interest in integrated optics for
ultra-high-speed optical signal processing and optical
 switching \cite{ultr} as well as in the generation and transmission of
nonclassical light.
In the view of these reasons it is certainly useful to make concrete efforts
to study the actual possibility to generate and to transfer squeezed light in this
system.

It is important to mention   various works  investigating
 similar systems, e.g  nonlinear asymmetric
directional couplers which consist of a linear waveguide and a nonlinear
waveguide operating by
second harmonic generation \cite{j1,{per2},{per5}} or
of a nonlinear waveguide with the core of Kerr-like medium \cite{kiy};
it has been shown that such systems can be used to reduce the necessary switching
power (for the switching of a signal controlled by a strong pump)
\cite{peter} as well as they are useful for constructing a band-pass
 power-filters or a band-reject power filter \cite{kiy}.
For a review of the role of quantum statistical
properties in nonlinear couplers, see \cite{prog}.

In the present work, we show that a nonlinear asymmetric directional coupler
composed of a linear waveguide and a nonlinear waveguide operating by
nondegenerate parametric amplification is an effective source of squeezed light.
This will be done as follows:
In  section 2 we describe the model under
discussion
and give the solution of the equations of motion.
In  section 3 we demonstrate the single-mode quasidistribution
functions.  Sections 4 and  5 are devoted to the photon-number
distribution and  phase distribution, respectively.
 Finally we  summarize the main conclusions in  section 6.

\section{ Equations of motion}
We consider a device (nonlinear asymmetric directional coupler)
as outline in Fig. 1.  It can be described by the
Hamiltonian

\begin{math}
{\displaystyle
\frac{\hat{H}}{\hbar}=[\sum_{j=1}^{3}\omega_{j}\hat{a}^{\dagger}_{j}\hat{a}_{j} ]+
i\lambda_{1}
\left\{
\hat{a}_{1}\hat{a}_{2}\exp[i(\omega_{1}+\omega_{2})t]-{\rm h.c.}\right\}
}\hfill
\end{math}

\begin{math}
{\displaystyle
+i\lambda_{2}
\left\{\hat{a}_{1}\hat{a}^{\dagger}_{3}\exp[i(\omega_{1}-\omega_{3})t]-{\rm
h.c.}\right\}
+i\lambda_{3}
\left\{\hat{a}_{2}\hat{a}^{\dagger}_{3}\exp[i(\omega_{2}-\omega_{3})t]-{\rm
h.c.}\right\},
}\hfill (1)
\end{math}

\noindent where
 $\hat{a}_{j} (\hat{a}^{\dagger}_{j}),\quad j=1,2,3$ are
 the annihilation (creation) operators designated to the signal, idler
 and linear modes, respectively,
$\lambda_{1}$ and $\lambda_{j}, \quad j=2,3$ are the
corresponding nonlinear and linear coupling constants;
$\omega_{j}$ are the natural  frequencies of oscillations
of the uncoupled modes and {\rm h.c.} is the Hermitian conjugate.
The linear exchange between the two waveguides
establishes through evanescent wave provided that
$\omega_{1}\simeq\omega_{2}\simeq \omega_{3}$. Moreover,  we treat the problem of propagation in the
Hamiltonian formalism neglecting dispersion. Thus if
all waves are propagating with the same velocity, time $t$ and space
$z$ relate by the velocity of propagation $v$, $z=vt$.
For further details of quantum description of propagation we refer the
reader to \cite{torn} (and references therein).
On the other hand, as we assume one-passage propagation, losses in the
 beams can  be neglected.
In general they can be described in the standard quantum
way in the form of interaction of light beams with reservoirs, as for
instance described in \cite{perin}. However, one can  note that
nonclassical properties of light beams degrade by the effect of the
reservoir, less by the damping which may be  by several orders weaker than
the nonlinear coupling, more by the influence of non-zero mean
numbers of reservoir oscillators.

\begin{picture}(70,70)(50,20)
\put (107,37){\vector(1,0){223}$\hat{a}_{1}(\frac{L}{v})$}

\put (107,60){\vector(1,0){223}$\hat{a}_{2}(\frac{L}{v})$}
\put (120,70){\makebox(0,0){$\hat{a}_{2}(0)$}}
\put (120,45){\makebox(0,0){$\hat{a}_{1}(0)$}}
\put (107,-20){\vector(1,0){223}$\hat{a}_{3}(\frac{L}{v})$}
\put (120,-10){\makebox(0,0){$\hat{a}_{3}(0)$}}

\put (300,-14){\vector(0,1){77}}
\put (200,0){\vector(0,1){40}}

\put (300,0){\vector(0,-1){20}}
\put (200,0){\vector(0,-1){20}}

\put (310,20){\makebox(0,0){$\lambda_{3}$}}
\put (210,20){\makebox(0,0){$\lambda_{2}$}}

\put (178,-50){\framebox(140,40){$\chi^{(1)}$}}
\put (178,30){\framebox(140,40){$\chi^{(2)}\quad\lambda_{1}$}}

\put (248,-100){\vector(1,0){70}}
\put (248,-100){\vector(-1,0){70}}
\put (248,-90){\makebox(0,0){$L$}}

\put (178,30){\line(0,-1){10}}
\put (178,10){\line(0,-1){10}}
\put (178,-10){\line(0,-1){10}}
\put (178,-30){\line(0,-1){10}}
\put (178,-50){\line(0,-1){10}}
\put (178,-70){\line(0,-1){10}}
\put (178,-90){\line(0,-1){10}}

\put (318,30){\line(0,-1){10}}
\put (318,10){\line(0,-1){10}}
\put (318,-10){\line(0,-1){10}}
\put (318,-30){\line(0,-1){10}}
\put (318,-50){\line(0,-1){10}}
\put (318,-70){\line(0,-1){10}}
\put (318,-90){\line(0,-1){10}}

\put (120,10){\line(1,0){20}}
\put (150,10){\line(1,0){20}}
\put (180,10){\line(1,0){20}}
\put (210,10){\line(1,0){20}}
\put (240,10){\line(1,0){20}}
\put (270,10){\line(1,0){20}}
\put (300,10){\line(1,0){20}}
\put (330,10){\line(1,0){20}}

\end{picture}

\vspace{1.7in}

{\noindent \it {\bf Fig.1}: Scheme of realization of interaction in (1) using
a nonlinear asymmetric directional coupler which is composed of two optical
waveguides
fabricated from first-order ($\chi^{(1)}$) and
second-order
 ($\chi^{(2)}$) materials, where $\chi$ designate
susceptibility. Signal mode 1 and idler mode 2 propagate in the first
waveguide
and  linear mode 3 in the second waveguide. The interaction between
the signal and idler modes
 is established by strong pump coherent light, which is  not indicated in
  figure, with the coupling constant $\lambda_{1}$.
The interactions between the linear mode  and the signal and idler
are established linearly with the  coupling constants $\lambda_{2}$
and $\lambda_{3}$, respectively.  The beams are described by the
photon annihilation operators as indicated; $z=vt$ is the interaction
length
and we assume that all beams have the same velocity $v$ and the
length of the waveguides is $L$.
Outgoing fields are examined as single or
compound
modes by means of homodyne detection to observe squeezing of vacuum
fluctuations, or by means of a set of photodetectors to measure photon
correlations, photon
antibunching and sub-Poissonian photon statistics in the standard ways.}

  The dynamics of the system is described by the Heisenberg
  equations of motion which, using the slowly varying forms
($\hat{a}_{j}=\hat{A}_{j}\exp(-i\omega_{j}t),\quad j=1,2,3$) to
eliminate
free oscillations so that only the slow dynamics due to the coupling
between modes appear explicitly,  read

\begin{math}
{\displaystyle
\frac{d\hat{A}_{1}}{dt}=-
\lambda_{1} \hat{A}_{2}^{\dagger}-\lambda_{2} \hat{A}_{3}  },\hfill
\end{math}

\begin{math}
{\displaystyle
\frac{d\hat{A}_{2}}{dt}=-\lambda_{1} \hat{A}_{1}^{\dagger}
-\lambda_{3} \hat{A}_{3}  },\hfill
\end{math}

\begin{math}
{\displaystyle
\frac{d\hat{A}_{3}}{dt}=\lambda_{2} \hat{A}_{1} +\lambda_{3} \hat{A}_{2}
 }.\hfill (2)
\end{math}

\noindent
These are  three equations  with their Hermitian conjugates
forming a closed system which can be solved easily by
the  Laplace transformation, restricting ourselves to the case
$\lambda_{2}=\lambda_{3}=\frac{\lambda_{1}}{\sqrt{2}}$, to avoid the complexity
in the calculations; this means that we consider stronger  nonlinearity
in the first waveguide compared with the linear exchange between
waveguides, thereby having the solution

\begin{math}
{\displaystyle
 \hat{A}_{1}(t)=\hat{a}_{1}(0)f_{1}(t) +\hat{a}^{\dagger}_{1}
(0)f_{2}(t)-\hat{a}_{2}(0)f_{3}(t)
}\hfill
\end{math}

\begin{math}
{\displaystyle
- \hat{a}_{2}^{\dagger}(0)f_{4}(t)-\hat{a}_{3}(0)f_{5}(t)-
\hat{a}^{\dagger}_{3}
(0)f_{6}(t) },\hfill
\end{math}

$ \hfill $

\begin{math}
{\displaystyle
 \hat{A}_{2}(t)=\hat{a}_{2}(0)g_{1}(t) +\hat{a}^{\dagger}_{2}
(0)g_{2}(t) -\hat{a}_{1}(0)g_{3}(t)
 }\hfill
\end{math}

\begin{math}
{\displaystyle
- \hat{a}_{1}^{\dagger}(0)g_{4}(t)-\hat{a}_{3}(0)g_{5}(t)-\hat{a}^{\dagger}_{3}
(0)g_{6}(t) },\hfill
\end{math}

$ \hfill $

\begin{math}
{\displaystyle
 \hat{A}_{3}(t)=\hat{a}_{3}(0)h_{1}(t) +\hat{a}^{\dagger}_{3}(0)h_{2}(t)
 +\hat{a}_{2}(0)h_{3}(t)}\hfill
 \end{math}

\begin{math}
{\displaystyle
+ \hat{a}_{2}^{
\dagger}(0)h_{4}(t)+\hat{a}_{1}(0)h_{5}(t)+\hat{a}^{\dagger}_{1}(0)h_{6}(
t) },\hfill (3)
\end{math}

 \noindent where  the time-dependent coefficients,
i.e.  $f_{j}(t),g_{j}(t),h_{j}(t)$,   including all information about
 the system  are

\begin{math}
{\displaystyle
f_{1,3}(t)=
\frac{1}{2} \left[
\cosh (\lambda_{1}t)
\pm \cosh (\frac{\lambda_{1}t}{2})\cos (\bar{k}t)
\pm\frac{1}{\sqrt{3}}
\sinh (\frac{\lambda_{1}t}{2})\sin (\bar{k}t)
\right],} \hfill
\end{math}

\begin{math}
{\displaystyle
f_{2,4}(t)=
\frac{1}{2} \left[
\sinh (\lambda_{1}t)
\mp\sinh (\frac{\lambda_{1}}{2}t)\cos (\bar{k}t)
\mp\frac{1}{\sqrt{3}}
\cosh (\frac{\lambda_{1}t}{2})\sin (\bar{k}t)
\right],} \hfill
\end{math}

\begin{math}
{\displaystyle
f_{5}(t)=
\sqrt{\frac{2}{3}}
\cosh (\frac{\lambda_{1}t}{2})\sin (\bar{k}t)
,\quad
f_{6}(t)=
-\sqrt{\frac{2}{3}}
\sinh (\frac{\lambda_{1}t}{2})\sin (\bar{k}t)
,} \hfill (4)
\end{math}

$\hfill$

\begin{math}
{\displaystyle
h_{1}(t)=
 \cosh (\frac{\lambda_{1}t}{2})\cos (\bar{k}t)
-\frac{1}{\sqrt{3}}
\sinh (\frac{\lambda_{1}t}{2})\sin (\bar{k}t)
,} \hfill
\end{math}

\begin{math}
{\displaystyle
h_{2}(t)=
- \sinh (\frac{\lambda_{1}t}{2})\cos (\bar{k}t)
+\frac{1}{\sqrt{3}}
\cosh (\frac{\lambda_{1}t}{2})\sin (\bar{k}t)
,} \hfill
\end{math}

\begin{math}
{\displaystyle
h_{3,5}(t)= \sqrt{\frac{2}{3}}
\cosh (\frac{\lambda_{1}t}{2})\sin (\bar{k}t)
,\quad
h_{4,6}(t)= -\sqrt{\frac{2}{3}}
\sinh (\frac{\lambda_{1}t}{2})\sin (\bar{k}t)
.} \hfill (5)
\end{math}

\noindent where $\bar{k}=\frac{\sqrt{3}}{2}\lambda_{1}$ and
 the expressions for $g_{j}(t)$  are the same  as those of
 $f_{j}(t)$.

In fact, the nature of the solution can show how the coupler does work.
To be more specific, the time-dependent coefficients  contain both
trigonometric and hyperbolic functions.
Consequently, the propagating beams inside the coupler
can be amplified as well as switched between waveguides, i.e. between
modes, in the course of time.

On the basis of the well known commutation rules for boson operators,
the following
relations can be proved for the time-dependent coefficients

\begin{math}
{\displaystyle
f^{2}_{1}(t)-f^{2}_{2}(t)+f^{2}_{3}(t)-f^{2}_{4}(t)+f^{2}_{5}(t)
-f^{2}_{6}(t)=1,} \hfill
\end{math}

\begin{math}
{\displaystyle
f_{1}(t)g_{4}(t)
-f_{2}(t)g_{3}(t)+f_{3}(t)g_{2}(t)-f_{4}(t)g_{1}(t)-f_{5}(t)g_{6}(t)
+f_{6}(t)g_{5}(t)=0,} \hfill
\end{math}

\begin{math}
{\displaystyle
f_{1}(t)g_{3}(t)
-f_{2}(t)g_{4}(t)+f_{3}(t)g_{1}(t)-f_{4}(t)g_{2}(t)-f_{5}(t)g_{5}(t)
+f_{6}(t)g_{6}(t)=0.} \hfill  (6)
\end{math}

\noindent The remaining relations  can be obtained from (6)
by means of the following transformations

\begin{math}
{\displaystyle
\Bigl(f_{1}(t),f_{2}(t),f_{3}(t)
,f_{4}(t),f_{5}(t),f_{6}(t)\Bigr)
} \hfill
\end{math}

\begin{math}
{\displaystyle
\longleftrightarrow
\Bigl(-g_{3}(t),-g_{4}(t),-g_{1}(t)
,-g_{2}(t),g_{5}(t),g_{6}(t)\Bigr)
} \hfill
\end{math}

\begin{math}
{\displaystyle
\longleftrightarrow
\Bigl(h_{5}(t),h_{6}(t),-h_{3}(t)
,-h_{4}(t),-h_{1}(t),-h_{2}(t)\Bigr).
} \hfill (7)
\end{math}

Based on the results of the present section,
we can study the quantum properties of the evolution of  different
modes in the model  when they are
initially prepared  in coherent states. This will be demonstrated in the following
sections.

\section{Quasiprobability functions}

Evaluation of various time-dependent mode observable is most
conveniently achieved with the aid of corresponding time-dependent
characteristic functions, their normal, antinormal and symmetric
forms, and the Fourier transforms of these characteristic
functions (quasiprobability functions). All of these are related
to the density matrix which provides a complete statistical
description of the system. There are three types of
quasiprobability functions: Wigner $W$-, Glauber $P$-, and Husimi
$Q$-functions. These functions could be used also as  crucial to
describe the nonclassical effects of the system, e.g. one can
employ the negative values of $W$-function, stretching of
$Q$-function and high singularities in $P$-function. Furthermore,
these functions are now accessible from measurements \cite{wig}.

In the following we consider  phase
space distributions  for the single-mode
 when all modes   are initially prepared in coherent  states before
 entering the coupler.

 We start from knowledge of the single-mode $s$-parameterized characteristic
 function which is suitable to describe the quantum statistics of light
 and for the $j$th mode it is defined by

\begin{math}
{\displaystyle C_{j}(\zeta,t,s)={\rm Tr} \left\{
\hat{\rho}(0)\exp \left[\zeta \hat{A}_{j}^{\dagger}(t)-
\zeta^{*} \hat{A}_{j}(t)+\frac{s}{2}|\zeta|^{2}\right] \right\},
 } \hfill (8)
\end{math}

\noindent where $\hat{\rho}(0)$ is the initial density matrix  for the
system  under consideration, $j$ takes on the values $1,2,3$ corresponding to
 considered mode
and $s$ takes on values $1, 0$ and $-1$ corresponding to
normally, symmetrically and antinormally ordered characteristic functions,
respectively.
Assume that the modes are initially uncorrelated and are described by the
 density operator

${\displaystyle \hat{\rho}(0)=\prod _{j=1}^{3}{\rm |\alpha
      \rangle_{j}}{\rm _{j}\langle \alpha |},}\hfill (9)$

\noindent where $|\alpha\rangle_{j}$ are  coherent states. So that
equations (8) and (9) lead to the single-mode  $s$-parameterized
characteristic function which for the signal mode takes the form

${\displaystyle
C_{1}(\zeta,s,t)=\exp\left\{ \frac{1
}{2}|\zeta|^{2}[s-|\eta _{1}(t)|]+
|\eta _{2}(t)| (\zeta^{2}+\zeta^{*2}
+[\bar{\alpha}_{1}^{*}(t)\zeta
-\bar{\alpha}_{1}(t)\zeta ^{*}]\right\} ,}\hfill (10)$

\noindent where $\bar{\alpha}_{1}(t)$ is the mean value of the
operator $\hat{A}_{1}(t)$ in the coherent state and $\eta
_{j}(t),\quad j=1,2$, are given by

${\displaystyle
\eta _{1}(t)=f^{2}_{1}(t)+f^{2}_{2}(t)+f^{2}_{3}(t)
+f^{2}_{4}(t)+f^{2}_{5}(t)+f^{2}_{6}(t)
,}\hfill $

${\displaystyle
\eta _{2}(t)=f_{1}(t)f_{2}(t) +f_{3}(t)f_{4}(t)+f_{5}(t)f_{6}(t).}\hfill
(11)$
\noindent

The single-mode $s$-parameterized quasiprobablity function is
defined as

\begin{math}
{\displaystyle
W_{j}(\beta,t,s)=\frac{1}{\pi^{2}}
\int C_{j}(\zeta,t,s)
\exp(\beta\zeta^{*}-\beta^{*}\zeta) d^{2}\zeta, }  \hfill (12)
\end{math}

\noindent  where $C_{j}(\zeta,t,s)$  is the $s$-parameterized
single-mode characteristic function and when $s=1,0,-1$ relation
(12) gives $P$-, $W$- and $Q$-functions, respectively.

Now the $s$-parameterized single-mode  quasiprobability  functions
of the signal mode, which can be derived from (10) and (12), are

${\displaystyle W_{1}(\beta,s,t)=\frac{2}{\pi\sqrt{[A_{+}(t)-s]
[A_{-}(t)-s]}}
}\hfill $

${\displaystyle
\times \exp\left\{ \frac{[l_{1}(t)-l^{*}_{1}(t)]^{2}}{2[A_{-}(t)-s]}
-\frac{[l_{1}(t)+l^{*}_{1}(t)]^{2}}{2[A_{+}(t)-s]}
\right\} ,}\hfill (13)$

\noindent where $l _{1}(t)=\bar{\alpha}_{1}(t)-\beta$,
 $\bar{\alpha}_{1}(t)$ has the same meaning as before and

${\displaystyle
A _{\pm}(t)=[f_{1}(t)\pm f_{2}(t)]^{2}+[f_{3}(t)\pm f_{4}(t)]^{2}
+[f_{5}(t)\pm f_{6}(t)]^{2};}
\hfill (14)$

\noindent the other expressions for the idler and linear modes can be
obtained from (13) and (14) by making use of the transformations (7); we
except here the case $s=1$, which will be discussed at the end of this
section.
Now let us start our investigation by demonstrating the behaviour
of the $W$-function, i.e. setting $s=0$ in (13).
In the language of mechanical analogy of a harmonic oscillator with dynamical
 conjugate variables $x$ and $y$, i.e $\beta=x+iy$, the $W$-function can be
 written in the form

${\displaystyle W_1(x,y,t)=\frac{1}{2\pi\sqrt{
\langle (\triangle\hat{X}(t))^{2}\rangle
\langle (\triangle\hat{Y}(t))^{2}\rangle
}}\exp\left\{- \frac{[x-\bar{\alpha}_{x}(t)]^{2}}{
2\langle (\triangle\hat{X}(t))^{2}\rangle}
-\frac{[y-\bar{\alpha}_{y}(t)]^{2}}{2
\langle (\triangle\hat{Y}(t))^{2}\rangle}
\right\}, }\hfill (15)$

\noindent where

${\displaystyle
\langle (\triangle\hat{X}(t))^{2}\rangle=\frac{A_{+}(t)}{4},\quad
\quad \quad
\langle (\triangle\hat{Y}(t))^{2}\rangle=\frac{A_{-}(t)}{4} ,}\hfill
(16)$

\noindent where
$\bar{\alpha}_{1}(t)=\bar{\alpha}_{x}(t)+i\bar{\alpha}_{y}(t)$;
$\langle (\triangle\hat{X}(t))^{2}\rangle$
 and $\langle (\triangle\hat{Y}(t))^{2}\rangle$ are   $x$- and
 $y$-quadrature variances, respectively.
As known these quadratures relate to the conjugate electric and magnetic
field operators of the electromagnetic field and may be used as a measure of
squeezing phenomenon. More illustratively,
the two quadrature operators of the $j$th single-mode are defined through
$\hat{X}(t)=\frac{1}{2}[\hat{A}_{j}(t)+\hat{A}_{j}^{\dagger}(t)], \quad
\hat{Y}(t)=\frac{1}{2i}[\hat{A}_{j}(t)-\hat{A}_{j}^{\dagger}(t)]$,
where
$[ \hat{X}(t),\hat{Y}(t)] =\frac{i}{2}$
and then the uncertainty-relation reads
$\langle (\triangle\hat{X}(t))^{2}\rangle
  \langle (\triangle\hat{Y}(t))^{2}\rangle \geq \frac{1}{16}$
where, e.g. $\langle (\triangle\hat{X}(t))^{2}\rangle
 = \langle (\hat{X}(t))^{2}\rangle
-\langle \hat{X}(t)\rangle^{2}$.
Therefore, we can say that the $j$th mode is  squeezed
 if $4\langle (\triangle \hat{X}(t))^{2}\rangle-1<0$  or
$4\langle (\triangle \hat{Y}(t))^{2}\rangle-1<0$.
Now from (14) and (16) together with (5)
one can easily prove that the quadrature variances of the linear mode
(single-mode squeezing) are

${\displaystyle
\langle (\triangle \hat{X}(t))^{2}\rangle=
\frac{1}{4}\left\{
\frac{4}{3}\sin ^{2}(\bar{k}t)+
[\cos (\bar{k}t)+
\frac{1}{\sqrt{3}}\sin (\bar{k}t)]^{2}\right\}
\exp (-\lambda_{1} t), }\hfill (17)$

$\hfill$

${\displaystyle
\langle (\triangle \hat{Y}(t))^{2}\rangle=
\frac{1}{4}\left\{
\frac{4}{3}\sin ^{2}(\bar{k}t)+
[\cos (\bar{k}t)-
\frac{1}{\sqrt{3}}\sin (\bar{k}t)]^{2}\right\}
\exp (\lambda_{1} t), }\hfill (18)$

\noindent where $\bar{k}$ has the same meaning as before.
Thus  squeezing can be achieved
in the linear waveguide in the $X$-quadrature.
Moreover, it is clear
that  squeezing values become more pronounced for a large interaction time,
i.e. for long length of the coupler $L$.
This behaviour shows that changing the power of the
linear interaction it is possible to transfer nonclassical properties,
which are generated in the nonlinear waveguide, to the linear signal mode.
For completeness, the uncertainty relation reads

${\displaystyle
\langle (\triangle\hat{X}(t))^{2}\rangle
\langle (\triangle\hat{Y}(t))^{2}\rangle=\frac{1}{16}[1
+\frac{16}{9}\sin ^{4}(\bar{k}t)]
.}\hfill (19)$

\noindent This formula reveals  that
the minimum-uncertainty relation  holds only when
$t=t_{s}=\frac{2m\pi}{\sqrt{3}\lambda_{1}}$ where $m$ is a positive
integer and we call $t_{s}$ a squeezed time.
In this
case the device provides the squeezed coherent light in the
linear mode with squeeze
parameter  $r=\frac{2m\pi}{\sqrt{3}}$.
It is important to mention that at $t=t_{s}$ the two waveguides become
completely independent (cf. equations (3)-(5)) and  the signal and idler modes
 can  display perfect two-mode
squeezing, which one can easily check with the help of two-mode quadrature
variances. This can be explained as follows: A portion of the energy
always remains within the guide into which the field was initially
injected. This energy grows in the course of time until $t=t_{s}$, when the
two guides are completely independent and hence the modes are trapped in
their own guides. So that  the self-interaction of the linear mode can
attribute the well known
single-mode squeezed light in the linear waveguide having its origin in
the nonlinear waveguide in which  the nonlinearity
 produces two-mode squeezing.
For later times $t\neq t_{s}$, the device generates
 single-mode squeezed light only in the linear waveguide where
the corresponding $W$-function may be broader than that of the squeezed
coherent states.   This situation is close
to that of a two-photon absorber (a two-photon absorption by a reservoir of
two-level atoms from a single mode of the electromagnetic field \cite{gil1})
 where a squeezed state, which is not a minimum-uncertainty state, has been
 generated \cite{gil}. The origin of squeezing of initially unsqueezed light
 interacting with two-photon absorbers is that the squeezing is generated
 by simple quantum superposition of states of light \cite{gil1}.

We proceed in our discussion by focusing our attention on the
$P$-function. It is known that the correspondence
 between quantum and classical theories can be established with the use
 of this function.
However, $P$-representation does not have all the
properties of a classical distribution function, especially for quantum
fields. To be more specific, light fields for which the $P$-representation
is not well-behaved distribution (in most processes involving an
interaction at least for some values of interaction time, including the
process under consideration) can exhibit  nonclassical features.
We have shown earlier that this model
is able to provide squeezed light and this should be reflected in the
 behaviour of the $P$-representation, i.e. setting $s=1$ in (13).
The significant example for this situation is the behaviour of the
linear mode. For this case, we can  show that the single-mode quadrature
 squeezing is established  provided that $A_{+}(t)-1<0$ and
 $A_{-}(t)-1>0$.  It is evident that the $P$-function is not
 well-behaved function in this case and this is the indication of
 the nonclassical fields.
  Indeed, if we look at the model under consideration as
   a competition of parametric
 processes, we can find that this result is in marked contrast
 to the most results occurred in the literature for three interacting
 modes
\cite{[12],{moll1}}, where the delta function of the initial
$P$-representation of coherent light  for a single mode
becomes well-behaved distribution  during the interaction.

In the following sections  we  use the phase space distribution
 functions to study the photon-number distribution and phase
 distribution for the system under discussion.

\section{ Photon-number distribution}

Photon-number distribution, i.e. the probability of finding  $n_{1}$
photons in the signal mode  at time $t$,
  can be obtained in the photodetection
process, and is determined by means of
 the relation \cite{[12]}

${\displaystyle
P(n_{1},t)=\langle n_{1}|\hat{\rho}_{1}(t)|n_{1}\rangle=
\int\frac{|\beta|^{2n_{1}}}{n_{1}!}
W(\beta,s=1,t)\exp (-|\beta|^{2}) d^{2}\beta;}
\hfill (20)$

\noindent substituting (13) into (20) and carrying out the integration,
we get the photon-number distribution of the signal mode as

${\displaystyle
P(n_{1},t)=\frac{2}{\sqrt{[A_{+}(t)+1][A_{-}(t)+1] }}
\exp \left\{ \frac{[\bar{\alpha}_{1}(t)-\bar{\alpha}^{*}_{1}(t)]^{2}}{
2[A_{-}(t)+1]}
-\frac{[\bar{\alpha}_{1}(t)+\bar{\alpha}^{*}_{1}(t)]^{2}}{2[A_{+}(t)+1]}
\right\} }\hfill$

${\displaystyle
\times \sum^{n_{1}}_{r=0}
\left[ \frac{A_{+}(t)-1}{A_{+}(t)+1}\right]^{r}
\left[ \frac{A_{-}(t)-1}{A_{-}(t)+1}\right]^{n_{1}-r}
}\hfill$

$ \hfill$

${\displaystyle
\times
{\rm L}^{-\frac{1}{2}}_{r}\left[
-\frac{(\bar{\alpha}_{1}(t)+ \bar{\alpha}^{*}_{1}(t) )^{2}}
{A^{2}_{+}(t)-1}\right]
 {\rm L}^{-\frac{1}{2}}_{n_{1}-r}\left[
\frac{(\bar{\alpha}_{1}(t)- \bar{\alpha}^{*}_{1}(t) )^{2}}
{A^{2}_{-}(t)-1}\right],}\hfill (21a)$

\noindent
where $L_{n}^{k}(x)$ is the associated Laguerre polynomials defined as %

${\displaystyle
L_{n}^{k}(x)=\sum_{m=0}^{n}\frac{\Gamma (n+k+1)
(-x)^{m}}{(n-m)!\Gamma
(m+k+1) m!},}\hfill (21b)$

\noindent
while $\Gamma $ is the Gamma function.

It is interesting to
compare this distribution with the corresponding Poisson distribution

${\displaystyle
P(n_{1},t)=\frac{\langle \hat{n}_{1}(t)\rangle ^{n_{1}}}{n_{1}!}\exp
(-\langle \hat{n}_{1}(t)\rangle ),}\hfill (22)$

\noindent
which corresponds
to fully coherent field with the same mean photon number
$\langle \hat{n}_{1}(t)\rangle$.
 The  expressions for the idler and linear modes should
 be obtained from (21a) using the transformations (7).
\setcounter{figure}{1}

\begin{figure}[h]%
    \includegraphics[width=8cm]{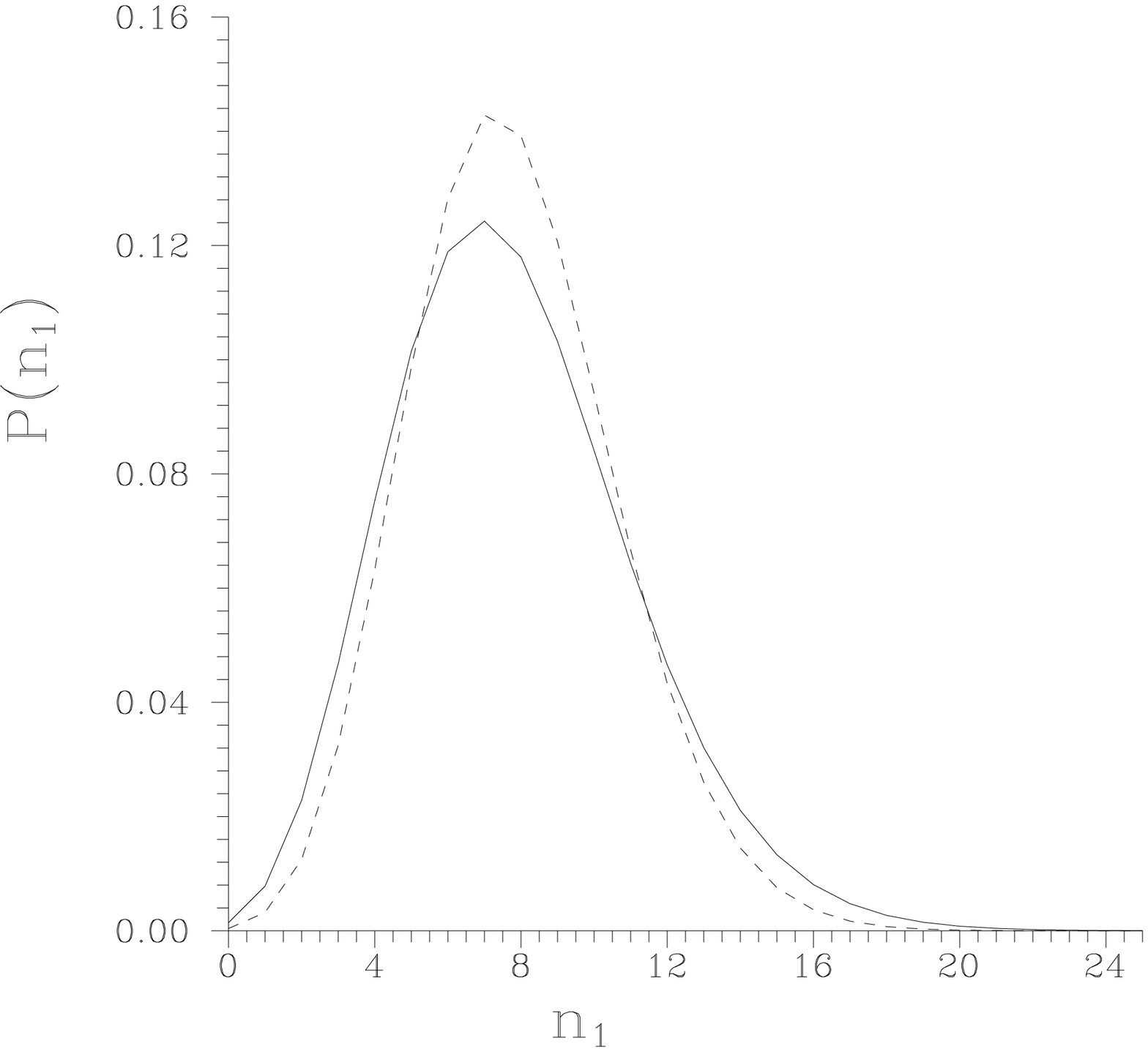}
   \caption{
Photon-number distribution $P(n_{1})$ for the signal mode  for
$\alpha_{j}=3\exp (i\frac{\pi}{3}), j=1,2,3, \quad t=1.5$ and
$\lambda_{1}=0.3$. The dashed curve is the Poisson photon-number
distribution.  }
  \label{fig1}
\end{figure}

In Fig. 2, we have plotted  $P(n_{1},t=1.5)$ against
$n_{1}$ for the signal mode  (solid curve) for $\alpha_{j}=
3\exp (i\frac{\pi}{3})$ and $\lambda_{1}=0.3$.
 For the sake of comparison, the corresponding
 photon distribution for coherent field, (22), is shown by dashed curve.
For all numerical calculations $\sum_{n_{1}=0}^{\infty}P(n_{1},t)=1$ with good
accuracy. We noted for the signal mode  that the photon-number distribution
exhibits always one-peak structure. From this figure we see that the
behaviour
is rather super-Poissonian as a result of quantum fluctuations because
 the solid curve is always broader than the corresponding dashed curve.
So that, in general the Poissonian light evolves in the nonlinear waveguide as
super-Poissonian light.
 On the other hand, as we have realized earlier from the behaviour  of
 $W$-function
that squeezed states can be generated. However, these states
are exhibiting oscillating photon-number distribution for certain values
of squeeze parameter. These oscillations are purely quantum effect without
classical analogue and they can be interpreted as interference in phase
 space \cite{sh} or in the framework of the generalized
 superposition of coherent fields and quantum noise \cite{bayer}.
Indeed, these oscillations  can be recognized for the  linear mode
where this mode  can display squeezed light
(see Fig. 3, for shown values of parameters).  From this figure we see  the
macroscopic oscillations of the photon-number distribution for squeezed light.
Further, comparison of Figs. 3a and 3b shows that the nonclassical
oscillations of $P(n_{3})$ are faster and more pronounced when $t$ increases.
 Moreover, we noted that these oscillations
appear only when the intensities of the input light are weak otherwise the
single-peak structure is dominant. The broader oscillation range
means that
$\langle (\triangle\hat{n}_{3}(t))^{2}\rangle>
\langle \hat{n}_{3}(t)\rangle$ and thus there are always super-Poissonian
statistics for the linear mode.
It is worthwhile mentioning that the case of Fig. 3a corresponds to
squeezed light with squeeze parameter
$r\simeq 3.6$.

\begin{figure}[h]%
  \centering
  \subfigure[]{\includegraphics[width=8cm]{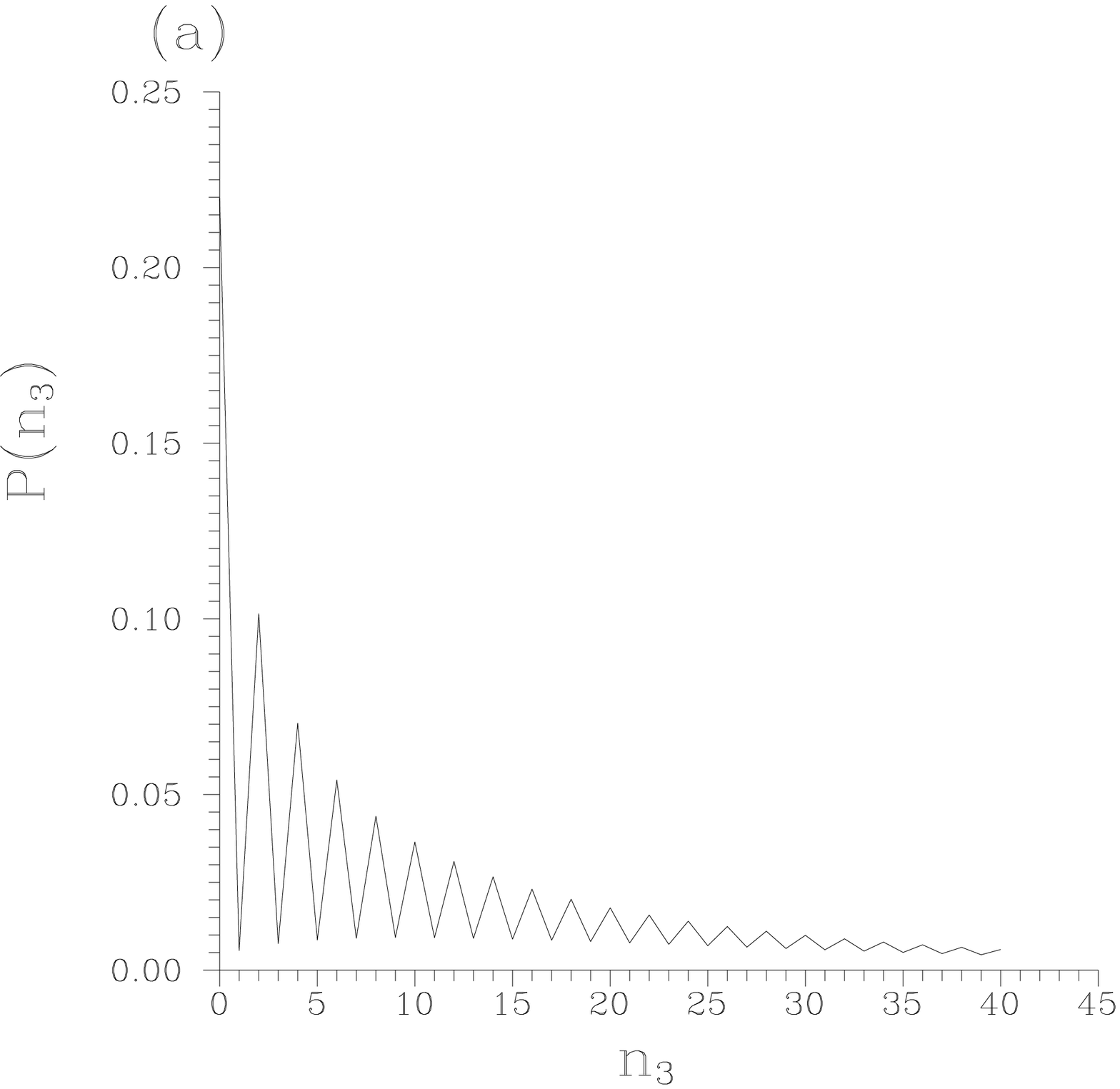}}
 \subfigure[]{\includegraphics[width=8cm]{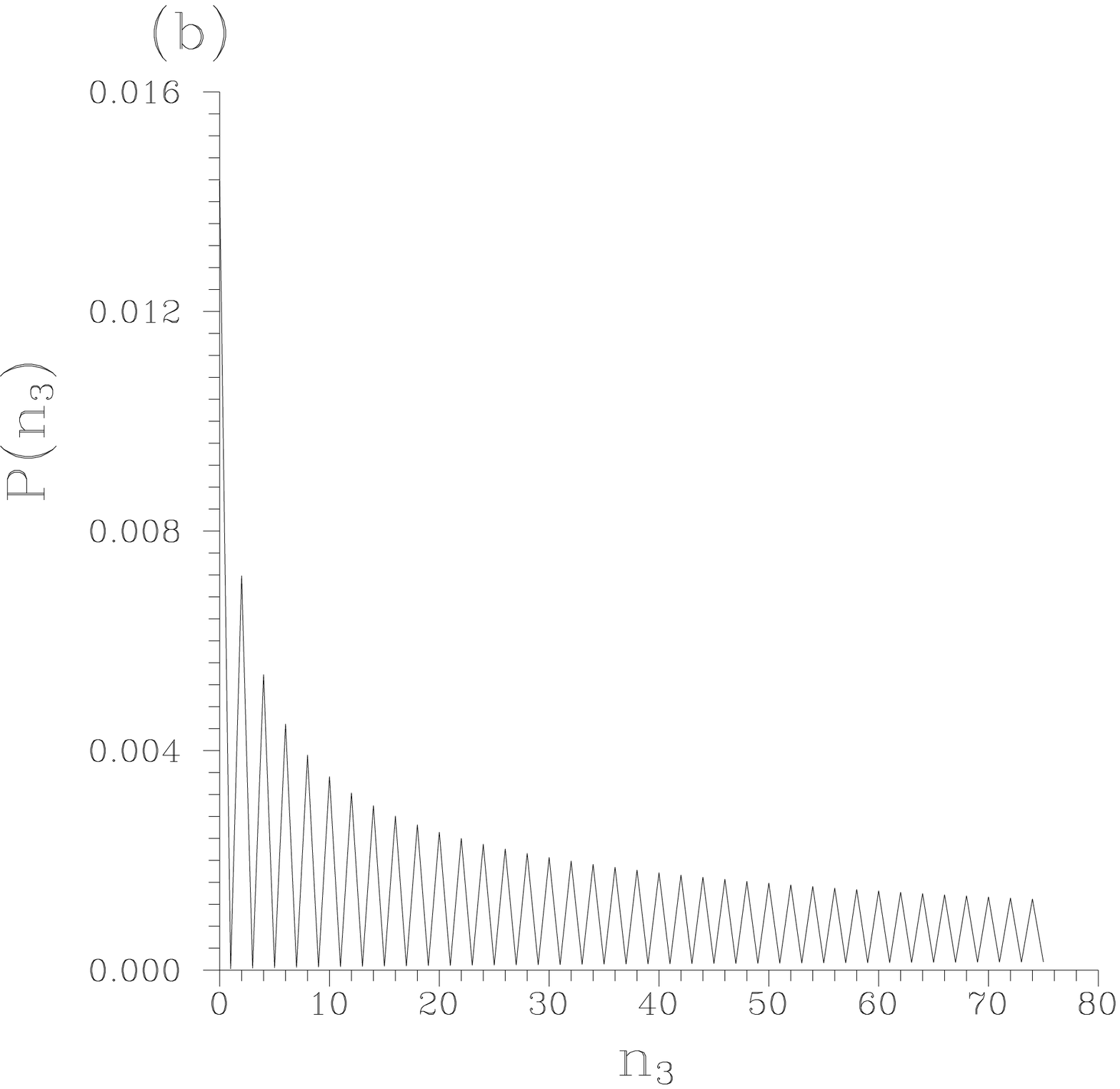}}
    \caption{
Photon-number distribution $P(n_{3})$ for the linear mode  when $
\alpha_{j}=0.5\exp(i\frac{\pi}{3}), j=1,2,3; \lambda_{1}=1$ and a)
$t=t_{s}$ ; b) $t=5$.  }
  \label{fig2}
\end{figure}

 In fact, the behaviour of the photon-number distribution in this
device is slightly different from that of the nonlinear asymmetric
coupler with strong classical stimulating light in the second
harmonic mode \cite{per2} where oscillatory behaviour as well as
sub-Poissonian statistics for specific modes in both linear and
nonlinear waveguide are exhibited.

\section {Phase distribution}

Nonlinear optical phenomena are sources of optical fields, the
statistical properties of which are changed  as
a result of the nonlinear interaction. Quantum phase properties are
among  those statistical properties which undergo nonlinear changes.
 The interest in the phase properties has been
motivated by experimental realization of optical homodyne tomography
\cite{tom} allowing quantum phase from the
measured field density matrix.

We  make use of the single-mode $Q$-function to investigate the phase
distribution for the system under discussion
 by integrating this function over the radial variable \cite{mir1}

${\displaystyle
P(\Theta,t)=\int_{0}^{\infty}W(\beta,s=-1,t) |\beta| d |\beta|.}\hfill
(23)$

\noindent It is worth mentioning that the authors of \cite{luis1,{luis2}} have
defined a Hermitian relative phase operator, which is in fact a polar
decomposition of the Stokes operators for the two-mode field.
In order to measure the single-mode phase distribution in this approach
we have to prepare one of the modes in a very intense state of well defined
phase and then the phase information of the other mode can be estimated
 \cite{luis1}. Recently, this approach has been experimentally
 demonstrated \cite{alex1}.  Actually,  this approach
gives different behaviour  than that  proposed from
quasiprobability distribution functions for weak fields
because of variety of possible definitions of quantum phase
\cite{mir1}. However,
they are in a good  agreement for strong fields  \cite{xper}.

We proceed, equations (13) and (23) lead to

${\displaystyle
P(\Theta,t)=\frac{1}{\pi C(t)\sqrt{A_{+}(t)A_{-}(t)}}
\exp
\left\{ \frac{[\bar{\alpha}_{1}(t)-\bar{\alpha}^{*}_{1}(t)]^{2}}{2[A_{-}(t)+1]}
-\frac{[\bar{\alpha}_{1}(t)+\bar{\alpha}^{*}_{1}(t)]^{2}}{2[A_{+}(t)+1]}\right\}
}\hfill $

${\displaystyle
\times \left\{ 1+\frac{B(t)}{2}\sqrt{\frac{\pi}{C(t)}}
\exp\left(\frac{B(t)^{2}}{4C(t)}\right)\left[1+{\rm erf}(\frac{B(t)}{2\sqrt{C(t)}})
\right]\right\},
}\hfill (24)$

\noindent where

${\displaystyle
C(t)=\frac{2}{A_{-}(t)A_{+}(t)}[A_{+}(t)\sin ^{2}\Theta+A_{-}(t)
\cos ^{2}\Theta],}\hfill $

$\hfill $

${\displaystyle
B(t)=\frac{2}{A_{-}(t)A_{+}(t)}\left\{ A_{-}(t)
[\bar{\alpha}_{1}(t)+\bar{\alpha}^{*}_{1}(t)]
\sin\Theta-iA_{+}(t)
[\bar{\alpha}_{1}(t)-\bar{\alpha}^{*}_{1}(t)]
\cos\Theta\right\},
}\hfill (25)$

\noindent and we have used the Gauss error function, which is defined by the
well-known formula

\begin{math}
{\displaystyle
{\rm erf}(x)=\frac{2 }{\sqrt{\pi}}\int^{x}_{0}\exp(-y^{2}) dy,
} \hfill  (26)
\end{math}

\noindent
taking into account that $\beta=|\beta|\exp(i\Theta)$.

It is clear that (24) is $2\pi$-periodic function and
the behaviour of this distribution  in this device
can be understood on the basis of the competition between the two-peak
structure for the vacuum states and the single-peak structure for coherent
states provided that the coherent amplitudes are real.
 Such behaviour is demonstrated
in Figs. 4a,b for the signal mode and in Figs.
5a,b for the linear mode, where the evolution of
the phase distribution (24) has been depicted
against the phase $\Theta$ and the time $t$.

\begin{figure}[h]%
  \centering
  \subfigure[]{\includegraphics[width=8cm]{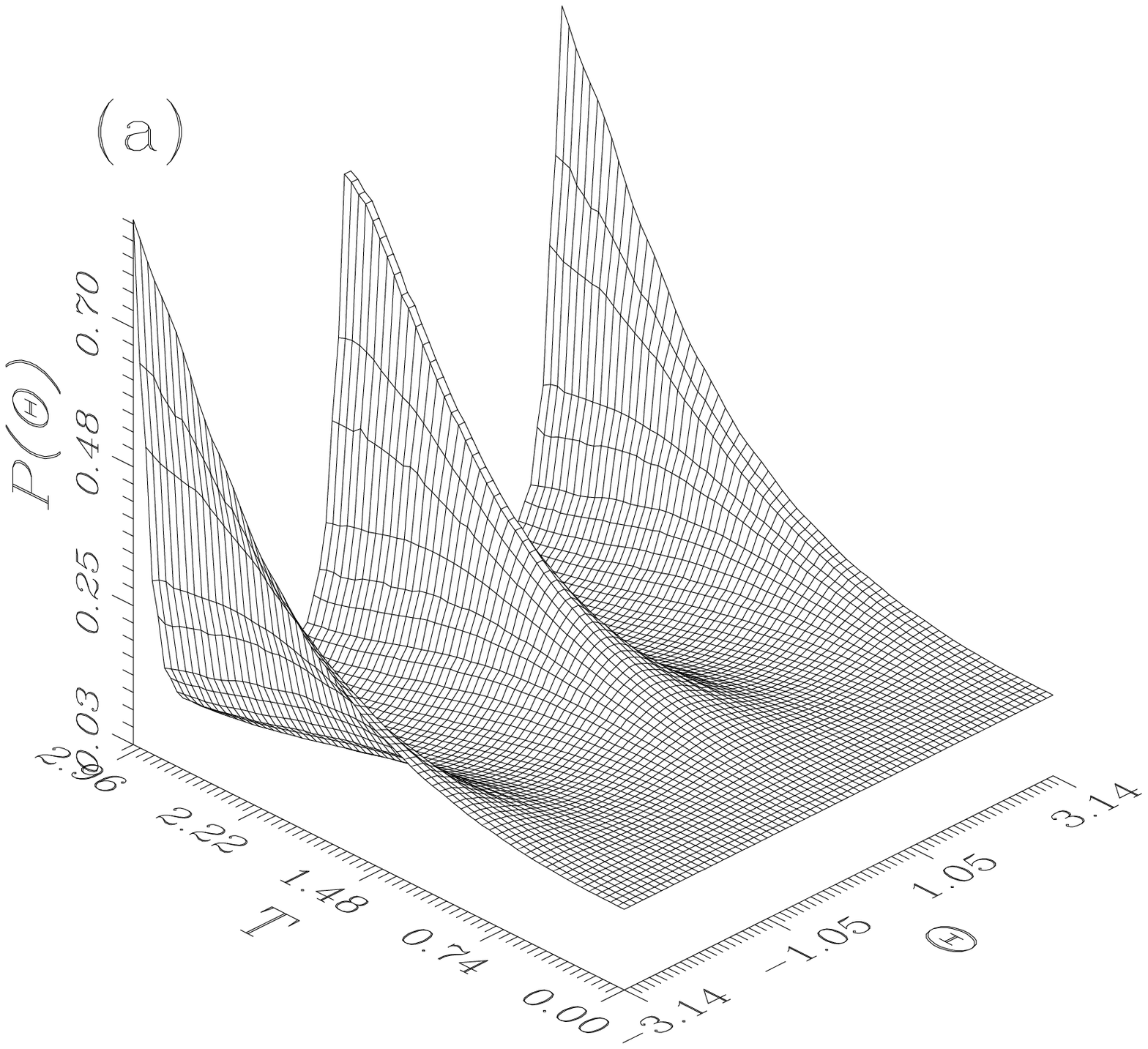}}
 \subfigure[]{\includegraphics[width=8cm]{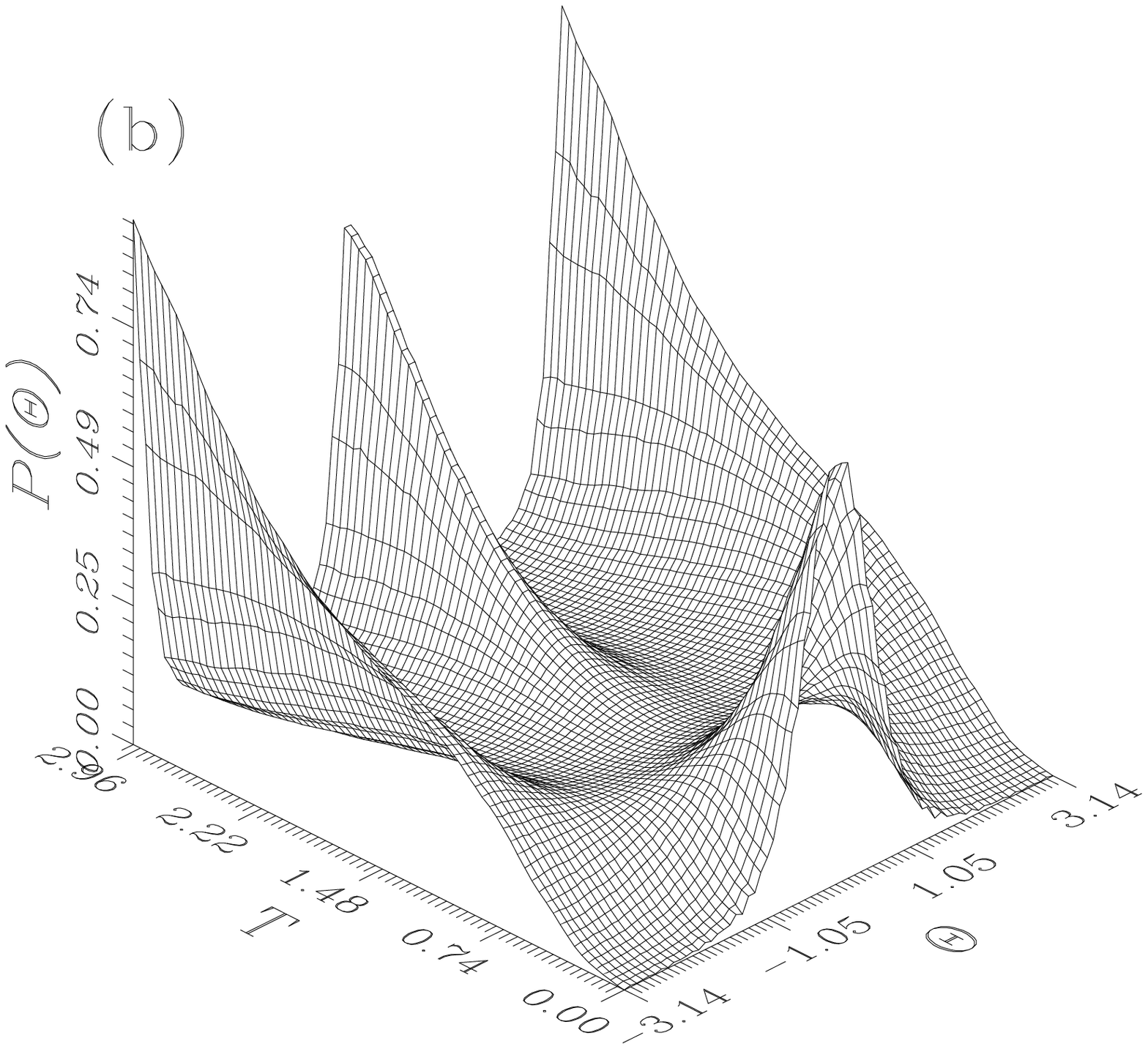}}
    \caption{
Phase distribution $P(\Theta,t)$ for the signal mode  against
$\Theta$ and $t$ for $\lambda_{1}=0.6$
 and a) $\alpha_{j}=0$; b) $\alpha_{j}=1,\quad j=1,2,3$. }
  \label{fig3}
\end{figure}

In figures a and b,
$\alpha_{j}=0$
and $1, \quad j=1,2,3$, respectively, for the shown  value of
coupling constant $\lambda_{1}$. In Fig. 4a, one can see the phase
evolution for the vacuum states in the signal mode showing two peak-structure
with one central peak at $\Theta=0$
and two wings as $\Theta\rightarrow \pm \pi$. In fact, this behaviour
 has been seen for some of superposition
states, e.g. for Yurke-Stoler states \cite{faisal1} and odd-binomial
states \cite{faisal2}.  In Fig. 4b, one can
observe that the initial peak for coherent state is attenuated in the
course of time while the two peaks for the vacuum states are created
 and amplified.
The well-known behaviour for the phase distribution of squeezed states
may be established when we
focus our attention on the behaviour of the linear mode, as expected (see Figs.
5a and b).

\begin{figure}[h]%
  \centering
  \subfigure[]{\includegraphics[width=8cm]{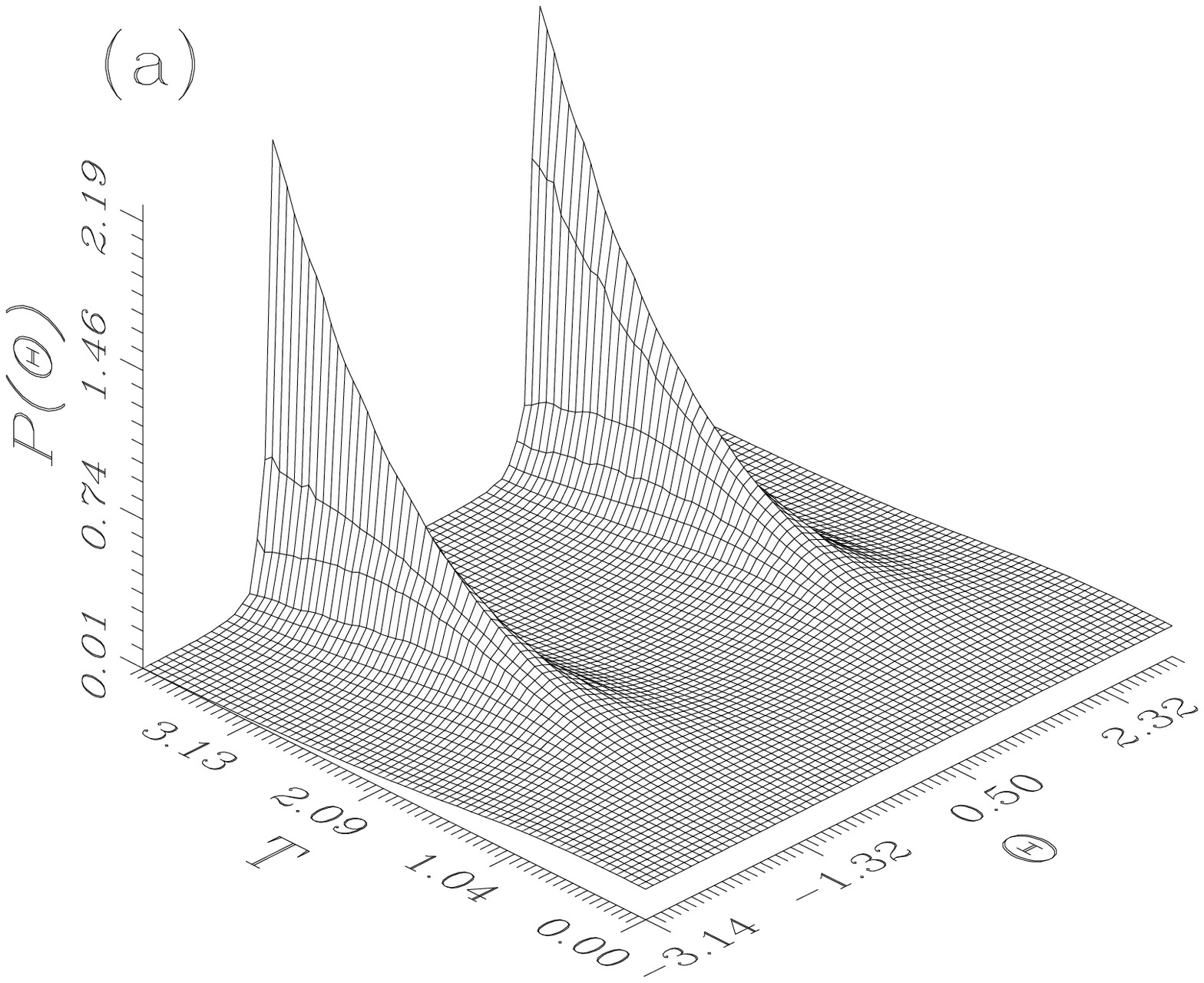}}
 \subfigure[]{\includegraphics[width=8cm]{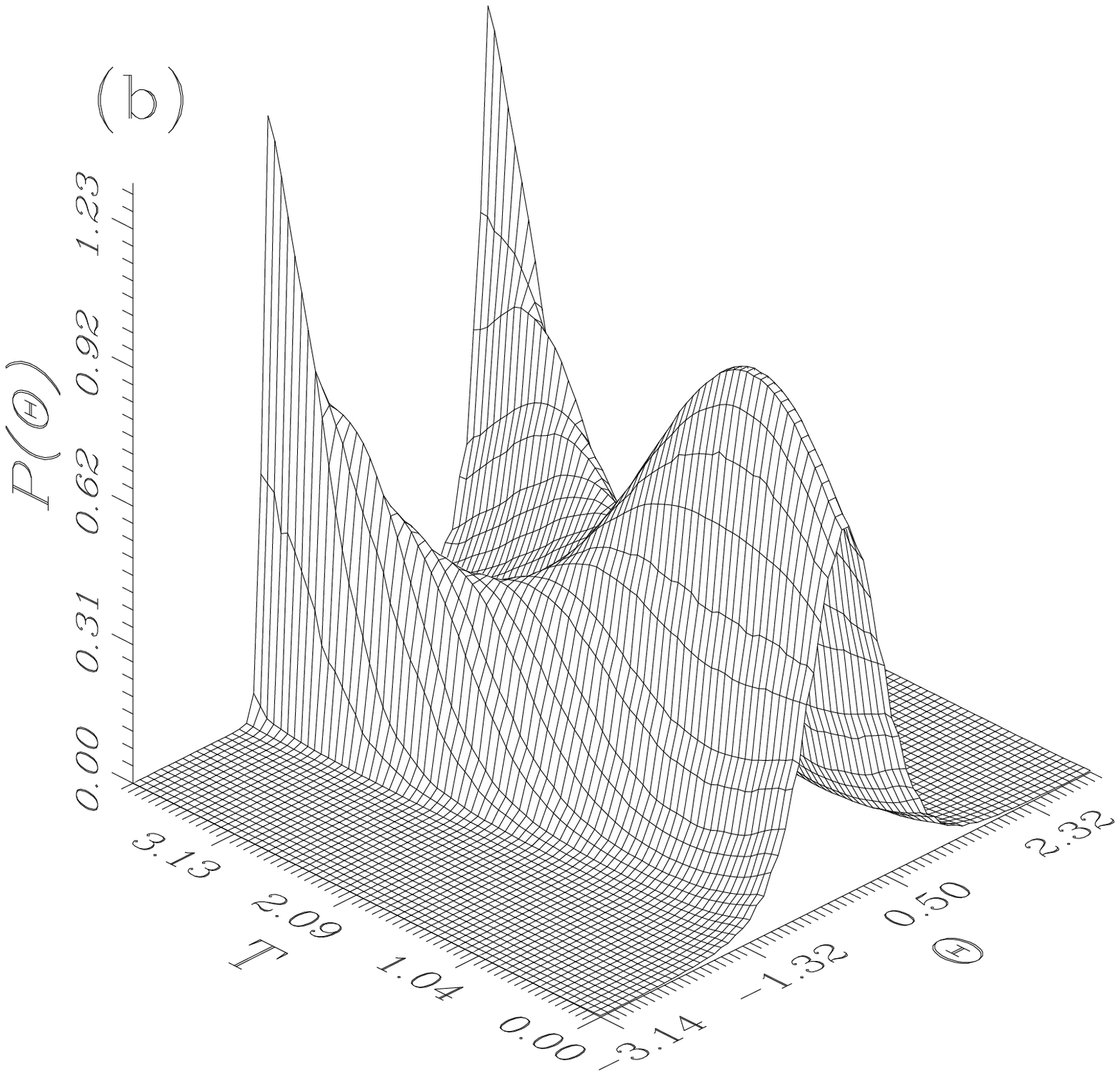}}
    \caption{
Phase distribution $P(\Theta,t)$ for the linear mode  against
$\Theta$ and $t$ for the same situation as in Fig. 4. }
  \label{fig4}
\end{figure}

 From figure 5a one can see that the two-peak structure
is typical for squeezed vacuum
states, i.e. we have two-peak structure for $\Theta=\pm\frac{\pi}{2}$
\cite{mir1}. Further,  the heights of these peaks
are $\frac{1}{2\pi}\sqrt{\frac{A_{-}(t)}{A_{+}(t)}}$ at any time $t\quad
(t>0)$. However, Fig. 5b displays the
well-known bifurcation shape for the phase distribution of squeezed field,
i.e. the distribution curve undergoes a transition from single- to a
double-peaked form with increasing time. Indeed this figure is quite similar
to that for squeezed coherent states \cite{schp}, and the distributions
differ in the behaviour of the initial peak, which is here amplified for
a while before splitting
into two peaks. This is connected with the significant influence of the
development by nonlinear effects and power transfers
between waveguides. It is worthwhile  mentioning that a similar behaviour
has been obtained for a contradirectional nonlinear asymmetric coupler
with strong stimulated coherent field in the second harmonic
waveguide \cite{per5}.

Finally, we would like to conclude this section by discussing
 the origin of such behaviour of the  phase distribution. This can be
easily
understood by analysing the function $P(\Theta,t)$ when $\alpha_{j}=0, \quad
j=1,2,3$.  In this case the formula (24) reduces to

${\displaystyle
P(\Theta,t)=\frac{1}{2\pi}\frac{\sqrt{A_{+}(t)A_{-}(t)}}{
A_{+}(t)\sin ^{2}\Theta+A_{-}(t)
\cos ^{2}\Theta}.}\hfill (27)$

\noindent This distribution function is double- or three-peak structure
according to the relation between $A_{+}(t)$ and $A_{-}(t)$.
 Let us restrict our discussion to the one of the modes which are
propagating in the
nonlinear waveguide, say, to the signal mode. For this mode, it can
easily be shown that
$A_{+}(t)>A_{-}(t)$ and consequently the formula (27) exhibits
three-peak structure with peaks for $\Theta=0,\pm \pi$. The heights of
these peaks at any time $t\neq 0$ are
$\frac{1}{2\pi}\sqrt{\frac{A_{+}(t)}{A_{-}(t)}}$. For non-zero
displacements $\alpha_{j}$, an additional factor of the form for coherent
state, but with a coherent amplitude $\bar{\alpha}_{1}(t)$ which is
the expectation
value of the signal mode operator in  coherent state, appears
in the distribution.
This factor is responsible for a peak at $\Theta=0$ related to coherent
component, which  competes with the
three-peak structure of vacuum to provide the previous behaviour of the
distribution. Similar argument can be adopted to explain the behaviour of the
linear mode, however, it should be borne in mind that in this case
$A_{+}(t)<A_{-}(t)$,  as we have shown earlier for squeezing phenomenon,
and hence the two-peak structure for the input vacuum case is dominant at
$\Theta=\pm\frac{\pi}{2}$.

\section{Conclusion}

In this article we have investigated quantum statistical properties of
 squeezed light generated  in a nonlinear asymmetric directional coupler
composed of a linear waveguide and a nonlinear waveguide operating by
nondegenerate parametric amplification.
The main difference between the  present system and the well
documented optical parametric oscillator (OPO) is the operating  mechanism of these devices.
More illustratively,  OPO is using a quadratic  nonlinear medium
 \cite{ex2}  in a resonator and it is
a practical device to generate squeezing; the system discussed
here is  richer by construction and by their properties because it
includes two propagating media (i.e. waveguides),  where one is
linear and the other is nonlinear and they  exchange energy
linearly  via evanescent waves. The system  considered here
consists of a mixture of nonlinear processes including parametric
generation and amplification (interaction coefficient
$\lambda_{1}$) and frequency conversion (interaction coefficients
$\lambda_{2}$ and $\lambda_{3}$). This can lead to a single mode
generation of nonclassical light.

Using the Heisenberg approach we have examined quantum statistics of
interacting modes, including the linear mode,
  propagating in the linear waveguide, in terms
of the quasiprobability functions, photon-number distribution and phase
distribution when
the modes are initially in coherent states.
From the behaviour of the Glauber  $P$-
and Wigner $W$-functions in the linear mode it follows
that the former is not
 well-behaved function when $t>0$, while,
  the latter displays the well-known behaviour for squeezed
light in which  one
 of the quadratures is amplified  and the other is attenuated.
In the photon-number distribution,
 the large scale macroscopic oscillations
related to squeezed light are established.
The phase distribution displays the
 bifurcation typical for squeezed field.
This behaviour demonstrates that changing the power of the
linear interaction, it is possible to transfer nonclassical properties,
which are generated in the nonlinear waveguide, to the linear signal
mode and to control them. Further, we have shown, in general, that the system is more suitable
for generation of squeezed light rather than sub-Poissonian light from  initial coherent light.

{\bf Acknowledgments}

We thank Prof. V. Pe\v{r}inov\'{a} and Dr. A. Luk\v{s} from
Department of Optics, Palack\'y University, Olomouc, Czech
Republic for the critical reading of the article. J. P. and F. A.
A. E-O. acknowledge the partial support from the Projects VS96028
and LN00A015 of Czech Ministry of Education. One of us (M. S. A.)
is grateful for the financial support from the Project Math
1418/19 of the Research Centre, College of Science, King Saud
University.

\end{document}